\documentclass[prl,aps,amsmath,amssymb,amsfonts,twocolumn,nofootinbib,superscriptaddress]{revtex4-1}

\usepackage{xcolor}
\usepackage{amsmath}
\usepackage{bbm}	
\usepackage{mathtools}
\usepackage{times}
\usepackage{hyperref}

\newcommand{\acom}[2]{\{#1,#2\}}

\newcommand{\ket}[1]{|#1\rangle}

\newcommand{\ketbra}[2]{|#1\rangle\langle#2|}

\newcommand{\sandwich}[3]{\langle#1|#2|#3\rangle}

\newcommand{\landau}{O}

\newcommand{\id}{\mathbbm{1}}
\newcommand{\tr}{\text{tr}}

\newcommand{\re}{Re}
\newcommand{\im}{Im}

\newcommand{\cc}{\mathbb{C}}
\newcommand{\rr}{\mathbb{R}}

\newcommand{\nn}{\mathbb{N}}

\newcommand{\hil}{\mathcal{H}}
\newcommand{\MPSMD}{\mathcal{M}_{D_\text{max}}}

\DeclareMathOperator*{\argmin}{arg\,min}

\definecolor{jens}{rgb}{.2,0.7,.9}

\definecolor{ors}{rgb}{.9,.0,.9}

\definecolor{libor}{rgb}{.5,.5,.1}

\begin{document}
\title{Fermionic orbital optimisation in tensor network states}
\author{C.\ Krumnow}
\affiliation{Dahlem Center for Complex Quantum Systems, Freie Universit{\"a}t Berlin, 14195 Berlin, Germany}
\author{L.\ Veis}
\affiliation{Strongly Correlated Systems ``Lend\"ulet'' Research Group, Wigner Research Centre for Physics, Hungarian Academy of Sciences, 1525 Budapest, Hungary}
\affiliation{J.\ Heyrovsky Institute of Physical Chemistry, Academy of Sciences of the Czech Republic, 18223 Prague, Czech Republic}
\author{\"O.\ Legeza}
\affiliation{Strongly Correlated Systems ``Lend\"ulet'' Research Group, Wigner Research Centre for Physics, Hungarian Academy of Sciences, 1525 Budapest, Hungary}
\author{J.\ Eisert}
\affiliation{Dahlem Center for Complex Quantum Systems, Freie Universit{\"a}t Berlin, 14195 Berlin, Germany}
\date{\today}

\begin{abstract}
Tensor network states and specifically matrix-product states have proven to be a powerful tool for simulating 
ground states of strongly correlated spin models. Recently, they have also been applied to interacting fermionic 
problems, specifically in the context of quantum chemistry. A new freedom arising in such non-local fermionic 
systems is the choice of orbitals, it being far from clear what choice of fermionic orbitals to make. In this work,
we propose a way to overcome this challenge. We suggest a method intertwining the optimisation over matrix 
product states with suitable fermionic Gaussian mode transformations. 
The described algorithm generalises basis changes in the spirit of the Hartree-Fock method
to matrix-product states, and provides a black box tool for basis optimisation in tensor network methods.
\end{abstract}
\maketitle

Capturing strongly correlated quantum systems is one of the major challenges of modern theoretical and computational physics. 
Recent years have seen a surge of interest in the development of potent numerical methods based on tensor networks 
to approximate ground states of interacting lattice models \cite{White1992,Schollwoeck2011,Verstraete-2008,Orus-2014,SchuchReview,EisertReview,AreaLaw},
building upon the success of the density-matrix renormalisation group  (DMRG) \cite{White1992}. It has become clear that such ideas are also applicable
to fermionic systems \cite{MERAF3,MERAF1,SchuchFermiPEPS}, and even to systems of quantum chemistry 
\cite{Fano1998,White1999,Mitrushenkov-2001,Chan2002,Legeza-2003a,Legeza-2003b,Moritz2005,Chan2011,Murg-2010a}, lacking the locality present in 
lattice models in condensed-matter systems. 
Such tools allow in principle to approximate the full configuration interaction
solution to good accuracy with reasonable effort, going 
in instances beyond conventional approaches to quantum chemistry, such as coupled cluster \cite{CC}, configuration interaction or 
density-functional theory \cite{KohnSham,Scheffler}, 
as convincingly shown by first implementations of DMRG algorithms in quantum chemistry (QC-DMRG)
\cite{Fano1998,White1999,Mitrushenkov-2001,Chan2002,Legeza-2003a}.

Yet, there is a new obstacle to be overcome: Tensor network methods have originally been tailored to capture local interactions, and consequently
ground states exhibiting short-range correlations and entanglement area laws \cite{AreaLaw}. Systems in quantum chemistry pose new challenges due to the
inherent long-ranged interactions, which are present no matter in what basis the systems are expressed. 
New questions hence arise concerning the optimal topology and \emph{physical (orbital) basis} 
used to construct the tensor network state 
\cite{Chan2002,Legeza-2003b,Mitrushenkov-2001,Moritz2005,Legeza-2003a,Rissler-2006,Chan2011,Murg-2010a,Fertitta-2014,Szalay-2015}.

In this work, we propose a novel approach towards making use of tensor network methods in quantum chemistry, by 
suggesting an adaptive scheme of updating basis transformations ``on the fly'' in conjunction with tensor network updates. In this way,
we bring together advantages of matrix product states -- which can capture strongly correlated states, but are tailored to short-ranged correlations and low
entanglement -- and fermionic Gaussian mode transformations -- for which entanglement is no obstacle, but non-Gaussian correlations are.
We hence go significantly beyond previous approaches towards optimising fermionic bases in tensor network approaches to quantum chemistry. 
Previous DMRG implementations in quantum chemistry allowing for an optimisation of the physical basis 
restrict the mode transformations to permutations and separate the optimisation over the basis and state such that multiple DMRG runs are necessary \cite{Barcza-2011}.
As a first attempt basis optimisations using a few transformations have been implemented for tree tensor networks, however, this has been found to be unstable \cite{Murg-2010a}.
Mixing fermionic orbitals from an active space -- the space considered
here -- with further ones from an additional external space has also been studied \cite{Nooijen,Chan,Luo-2010}. 
In these approaches orbital transformations are carried out again between
different DMRG runs. In contrast to this we perform the mode transformations within the active space
in parallel to the state optimisation and directly optimise the
entanglement structure of the tensor network.

We focus on matrix-product states, but explain in what way the idea is generally applicable. We also discuss the role of symmetries
and the geometry of the problem at hand. 
The basis optimisation is incorporated into the standard two-site QC-DMRG and can be added to existing implementation without increasing the computational costs of the DMRG. 
The resulting scheme can be used in parallel to a ground state search or as a pre-processing step in which the physical basis is optimised in a first phase restricting the bond dimension of the MPS used to medium values and calculating the final ground state in the optimised basis with higher accuracy. 

{\it System.}
In this work, we are concerned with 
strongly correlated interacting fermionic models with a finite number of relevant modes as they appear in the quantum chemistry
context.  In second quantised form the  Hamitonian takes the form
	\begin{equation} H = \sum\limits_{i,j=1}^{np} t_{i,j} c_i^\dag c_j + \sum\limits_{i,j,k,l=1}^{np} v_{i,j,k,l}c_i^\dag c_j^\dag c_l c_k, \label{eq:Hamiltonian}\end{equation} 
where $c_j$ is a fermionic annihilation operator associated to the mode labeled $j$ satisfying the canonical anti-commutation relations $\acom{c_i}{c_j}=0$ and $\acom{c_i^\dag}{c_j}=\delta_{i,j}$ and the coupling $t$ and $v$ are such that $H$ is Hermitian. 
$p$ denotes the number of different fermion species present for each of the $n$ orbitals, e.g.~spin up and down electrons. 
The one particle modes form the basis of single particle Hilbert space $\hil_{np}$. Any 
fermionic state will be an element of the fermionic Fock space $\mathcal{F}=\bigoplus_{k=0}^{np} \bigwedge^k \hil_{np}$, 
where $\wedge$ denotes the exterior product and $\wedge^0\hil_{np}=\cc$, with a basis formed of all Slater determinants $\ket{x}$, where $x\in \{0,1\}^{np}$, of the initial single particle modes.
We refer to this basis as the physical basis.
The Jordan Wigner transformation establishes an isomorphism between $\mathcal{F}$ and the Hilbert space of $n$ qudits $\hil_d^{\otimes n}=\cc^{d^n}$ with $p=\log_2d$.
By choosing any ordering of the orbitals such systems can be viewed as one-dimensional lattices of $n$ sites
with long-range interactions.
 
{\it MPS and general idea.}
For a one-dimensional quantum lattice with $n$ sites, where each site is described by a $d-$dimensional Hilbert space $\hil_d$, a matrix product state (MPS)
vector takes the general form
	\begin{equation} \ket{\psi} = \sum\limits_{\alpha_1,\ldots,\alpha_n=1}^d A^{\alpha_1}_{[1]}\ldots A^{\alpha_n}_{[n]}\ket{\alpha_1}\otimes\ldots\otimes \ket{\alpha_n},\label{eq:DefinitionMPS}\end{equation}
where $A_{[m]}^{\alpha_m}\in \cc^{D_{m-1}\times D_m}$ and $\{\ket{\alpha}\}$ form a basis of $\mathcal{H}^d$ and $D_0=1=D_n$. 
If the \emph{bond dimension} $D\in \nn$ is allowed to vary arbitrarily over different sites, every quantum state of the lattice can be written as in Eq.\ \eqref{eq:DefinitionMPS} \cite{Vidal2003}. 
Restricting the maximal bond dimension along the chain to a fixed value $D_\text{max}$ creates the sub-manifold $\MPSMD$ of the full state space.
Approximations of the ground state of a given Hamiltonian within this sub-manifold can be found using the density matrix renormalisation group algorithm (DMRG) which, as an alternating least square method, optimises the entries of the MPS tensors $(A_{[m]})_m$ iteratively \cite{Ostlund-1995,Verstraete-2004a,Verstraete-2004b,Szalay-2015}.

The freedom one has in this construction is a redefinition of the fermionic modes by a linear transformation.
Linear transformations of a set of fermionic annihilation operators $\{c_i\}$ to a new set $\{d_i\}$
satisfying the canonical anti-commutation relations
are captured by $c_i = \sum_{j=1}^{np} U_{i,j} d_j,$
with a unitary mode-transformation $U\in U(np)$.
This change of the single particle modes induces a transformation of the physical basis of $\mathcal{F}$. 
Under this change of basis a fermionic state vector $\ket{\psi(\id)}$ transforms to $\ket{\psi(U)}=G(U)\ket{\psi(\id)}$ with the Gaussian unitary 
transformation $G(U)=\exp{[\sum_{i,j} (\ln U^\dag)_{i,j} c^\dag_i c_j]}$ acting in Fock space.
The transformation on $\cc^{d^n}$ induced from the Jordan Wigner transformation is given by $g(U)=\bigoplus_{k=0}^{np}\bigwedge^k U^\dag$ where $\bigwedge^0 U^\dag=1$.

\begin{figure}
	\includegraphics{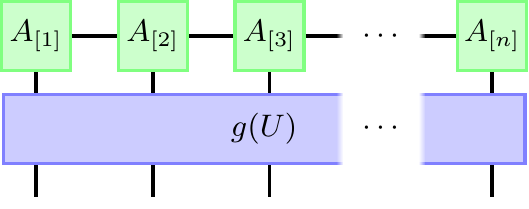}
	\caption{Illustration of the general ansatz-class of an MPS with varying physical basis, where $g(U)$ is a Gaussian transformations defined by a mode transformation $U\in U(np)$ as described in the main text.}
	\label{fig:MPSGU}
\end{figure}

We now turn to describing ground states of fermionic Hamiltonians with 
MPS expressed in a given basis, where the approximatability of the states strongly depends on the 
choice of basis \cite{Rissler-2006,Fertitta-2014}. 
Specifically, denoting the Hamiltonian written in terms of the transformed modes by $H(U) = G(U)^\dag H G(U)$, we are interested in the solutions of
	\begin{equation} (U_\text{opt},\ket{\psi_\text{opt}})=\argmin\limits_{U\in U(np),\ket{\psi}\in\MPSMD} \sandwich{\psi}{H(U)}{\psi}. \label{eq:OptimizationH}\end{equation}
Note that the Hartree-Fock method is readily included in Eq.\ \eqref{eq:OptimizationH} by the case $D_\text{max}=1$, when 
$\ket{\psi}$ is restricted to be a Slater determinant.
Identifying the optimal or close-to optimal basis for a general Hamiltonian and $D_{\text{max}}$ in the sense of Eq.\ \eqref{eq:OptimizationH} would provide a deeper understanding of the entanglement structure of ground states appearing in quantum chemistry, but since this is a non-convex problem, approximate solutions are accessible only.
Here, we take an approach that
iteratively finds close to optimal solutions numerically by optimising over the ansatz-class depicted in Fig.~\ref{fig:MPSGU}.

{\it Compositions of local mode transformations.}
In order to calculate approximations to the solutions of Eq.\ \eqref{eq:OptimizationH} and avoiding stability and performance issues of a direct global optimisation, we perform successive local mode transformations in parallel to a two-site QC-DMRG and use a few additional global reorderings of the orbitals as in Refs.\ \cite{Barcza-2011,Fertitta-2014} to leave local minima during the optimisation-process. 
Given a state vector $\ket{\psi}$, a site-index $m\in[n-1]$ and a cost function $f_m$ which will be discussed below we solve
	\begin{equation} U_\text{opt}^\text{loc} = \argmin\limits_{U\in V}f_m\bigl(\ket{\psi(\id_{pm}\oplus U\oplus \id_{pn-pm-2p})}\bigr),\label{eq:OptimizationLocalPsi}\end{equation}
with $\id_k$ denoting the $k-$dimensional identity matrix and $V\subset U(2p)$ needs to be chosen depending on the symmetries of the system.
The global basis change is then composed of local unitaries which are solutions of \eqref{eq:OptimizationLocalPsi} for different $m$ and act non-trivially on overlapping areas of the lattice and intermediate global reorderings of the lattice-sites.

The cost function is chosen according to the following paradigm.
The bond dimension needed for a bipartition of the system to approximate a state up to a predefined accuracy can be upper bounded using the R{\'e}nyi entropies $S^\alpha(\rho_\text{red})$ of the 
reduced state for $\alpha < 1$ \cite{Verstraete2006}, where $S^\alpha(\rho)=\log \tr \rho^\alpha/(1-\alpha)$. 
We therefore iteratively minimise the $S^{\frac{1}{2}}$ entropy over the chosen bipartition by using the cost function $f_m^{(1)}(\ket{\psi}) = ||\Sigma^m_\psi||_1$ where $\Sigma^m_\psi$ denotes the Schmidt spectrum of $\ket{\psi}$ for a bipartiting cut between sites $m$ and $m+1$.
With increasing dimension of $V$, which growth with the number of species per orbital $p$, and bond dimension $D_\text{max}$ the optimisation of $f^{(1)}_m$ becomes slow.
Efficiency can be gained by minimising $f^{(4)}_m=-||\Sigma_\psi^m||_4^4$ of which we can calculate the gradient $\nabla_{U_{ij}}f^{(4)}_m(\ket{\psi(\id_{pm}\oplus U \oplus \id_{pn-pm-2p})})$ analytically and efficiently in the bond dimension as shown in the appendix.
The optimisation of $S^2$ will not lead to certified bounds on the required bond dimension, but will favour stronger decays in the Schmidt spectrum similar as the minimisation of $S^\frac{1}{2}$.

The results presented here have been obtained by optimising the one norm of the Schmidt spectrum, $f_m^{(1)}$. 
The optimisation of $f^{(4)}_m$ can be applied in if $V$ has a higher dimension; which appear for $p>1$ if the system lacks symmetries.
Both the choice of the cost function and symmetries influence the choice of $V$ as argued in the following.

{\it Optimisation set.}
In the presence of symmetries, choosing a physical basis which can be labeled by good quantum numbers decouples different symmetry sectors in the coefficient tensors of the Hamitonian and the MPS and allows for more efficient computations.
Only mode transformations which commute with the generators of the symmetry transformations will preserve the structure imposed by the symmetry.

In general QC-DMRG algorithms only exploit a subgroup of the full symmetry group of a specific Hamiltonian, such as conservation of the number of particles, spin reflection symmetries, Abelian point group symmetries or a $SU(2)$ spin rotation symmetry \cite{White1992,Legeza-2003b,McCulloch-2007,Toth-2008,Sharma-2012a,Wouters-2012} (see also Ref.\ \cite{Szalay-2015}).
Here, we consider for the states the case of particle number conservation of each species, which is an Abelian symmetry and allows for an easy implementation of symmetric MPS \cite{Singh2011} and want the local mode transformations to respect the $SU(2)$ symmetry of the considered systems.
The admissible transformations in this case are of the form $U= U_n^{\oplus p}$ with $U_n\in U(n)$ acting on one species of fermions.

The cost functions $f_m$ chosen above depend only upon the Schmidt-spectrum of the state for a cut between sites $m$ and $m+1$ and are therefore insensitive to mode transformations of the form $U_m\oplus U_{m+1}$ with $U_q\in U(p)$ acting only on the modes associated to the lattice site $q$. 
To obtain a non-redundant parametrisation of the unitaries used in the optimisation in Eq.\ \eqref{eq:OptimizationLocalPsi} we restrict it to the set of left cosets $U(2p)/ U(p)\times U(p)$ which is isomorphic to the Grassmann manifold $G(2p,p)$. 
Efficient implementations of optimisation algorithms such as the conjugate gradient method within Grassmann manifolds using $2p^2$ parameters are described in Refs.\ \cite{Edelman1998,Manton2002}.
If we restrict ourselves to mode transformations which preserve the $SU(2)$-symmetry, the relevant mode transformations are parametrised by $G(2,1)$, leaving $2$ optimisation parameters in each step.
Focussing on this case here with medium values for $D_\text{max}$, we can obtain $U_\text{opt}^\text{loc}$ of $f^{(1)}_m$ by using gradient free schemes such as the Nelder-Mead method, due to the small number of parameters.

{\it Algorithm.}
Combined with an approximation of the ground state of a given Hamiltonian, local mode transformations naturally extend a two-site DMRG. 
A single two-site DMRG step results in a blocked tensor $A_{[m,m+1]}\in \cc^{d^2\times D_{m-1}\times D_{m+1}}$. 
In the generic case, restoring the MPS format in Eq.\ \eqref{eq:DefinitionMPS} by decomposing the blocked tensor into local tensors $A_{[m]}^\prime\in \cc^{d\times D_{m-1}\times D^\prime_m}$ and $A_{[m+1]}^\prime\in \cc^{d\times D^\prime_m \times D_{m+1}}$ will lead to $D^\prime_{m}>D_\text{max}$ such that the found state needs to be projected into $\MPSMD$ by discarding the $D_\text{max}-D_m^\prime$ smallest values of the resulting Schmidt spectrum $\Sigma_\psi^m$. 
The projection yields an truncation error $\epsilon_{t} = \sum_i \sigma_i^2$, where $\sigma_i\in\Sigma_\psi^m$ are the discarded singular values. 
If we allow for a local mode transformation before the blocked tensor is decomposed, the truncation error can be reduced.

\begin{table}[t]
\caption{Two site DMRG with adaptive mode transformations.}
\label{tbl:algorithm}

{\parindent0pt\flushleft
1\quad {\bf iterate} over neighbouring sites $m\in[n-1]$:

2\qquad get blocked tensor $A^{\alpha,\beta}_{[m,m+1]}$ (e.g.~from two-site DMRG)

3\qquad calculate (local) minimum $U_\text{opt}^\text{loc}$ of $f_m(\ket{\psi(\id\oplus U \oplus \id)})$

4\qquad {\bf if} $f_m\ket{\psi(\id\oplus U_\text{opt}^\text{loc}\oplus \id)}<f_m(\ket{\psi(\id)})$:

5\quad\qquad transform $\ket{\psi}$ with $U$ by updating\\\quad\qquad\hphantom{5}$A_{[m,m+1]}^{\alpha,\beta} = \sum_{\alpha^\prime,\beta^\prime=1}^dg(U_\text{opt}^\text{loc})_{(\alpha,\beta),(\alpha^\prime,\beta^\prime)}A_{[m,m+1]}^{\alpha^\prime,\beta^\prime}$ and\\\quad\qquad\hphantom{5}transform relevant operators with $U^\dag$

6\qquad calculate $A_{[m]}$, $A_{[m+1]}$ by decomposing $A_{[m,m+1]}$ with\\\qquad\hphantom{6}truncation and update MPS with new tensors\\
}
\end{table}

Using the gauge-invariance of MPS, we bring the MPS to a mixed normalised form, i.e. matrices of sites $q<m$ are left-normalised whereas matrices associated to sites $q>m+1$ are right-normalised \cite{Garcia2007,Schollwoeck2011}.
We can then calculate $\Sigma_\psi^m$ from the blocked tensor $A_{[m,m+1]}$. 
We optimise the basis by solving Eq.\ \eqref{eq:OptimizationLocalPsi} while keeping expectation values of the state, such as the energy, constant by using 
$\sandwich{\psi}{H}{\psi} = \sandwich{\psi(U)}{H(U^\dag)}{\psi(U)}$ with $U = \id_{pm}\oplus U_\text{opt}^\text{loc}\oplus \id_{pn-pm-2p}$.
As the mode transformation acts non-trivially only on sites $m$, $m+1$ the transformed state vector $\ket{\psi}$ can be represented by
\begin{eqnarray}
	A_{[k]}(U)&=&A_{[k]}(\id),\,\,k\in[n]\backslash\{m,m+1\},\\
	A_{[m,m+1]}^{\alpha,\beta}(U)&=&\sum_{\alpha^\prime,\beta^\prime}g(U)_{(\alpha,\beta),(\alpha^\prime,\beta^\prime)}
	A_{[m,m+1]}^{\alpha^\prime,\beta^\prime}(\id).
\end{eqnarray}	 
Operators such as the Hamiltonian can be transformed efficiently using their second quantised representation.
For an operator $O = \sum_{i_1,\ldots,i_s,j_1,\ldots,j_s=1}^{np} o_{i_1,\ldots,i_s,j_1,\ldots,j_r}c^\dag_{i_1}\ldots c^\dag_{i_s}c_{j_t}\ldots c_{j_1}$ with $o=o(\id)$, the coefficients transform under a mode-transformation according to $o(U) = (U^\dag)^{\otimes s} o(\id) U^{\otimes r}$.
As most operators of interest, e.g., the Hamiltonian, contain terms with small $s$ and $r$ those transformations can be implemented efficiently; with cost scaling as $\landau((np)^{s+r-1})$ for local transformations. 

\begin{figure}[t]
	\includegraphics[width=0.5\textwidth]{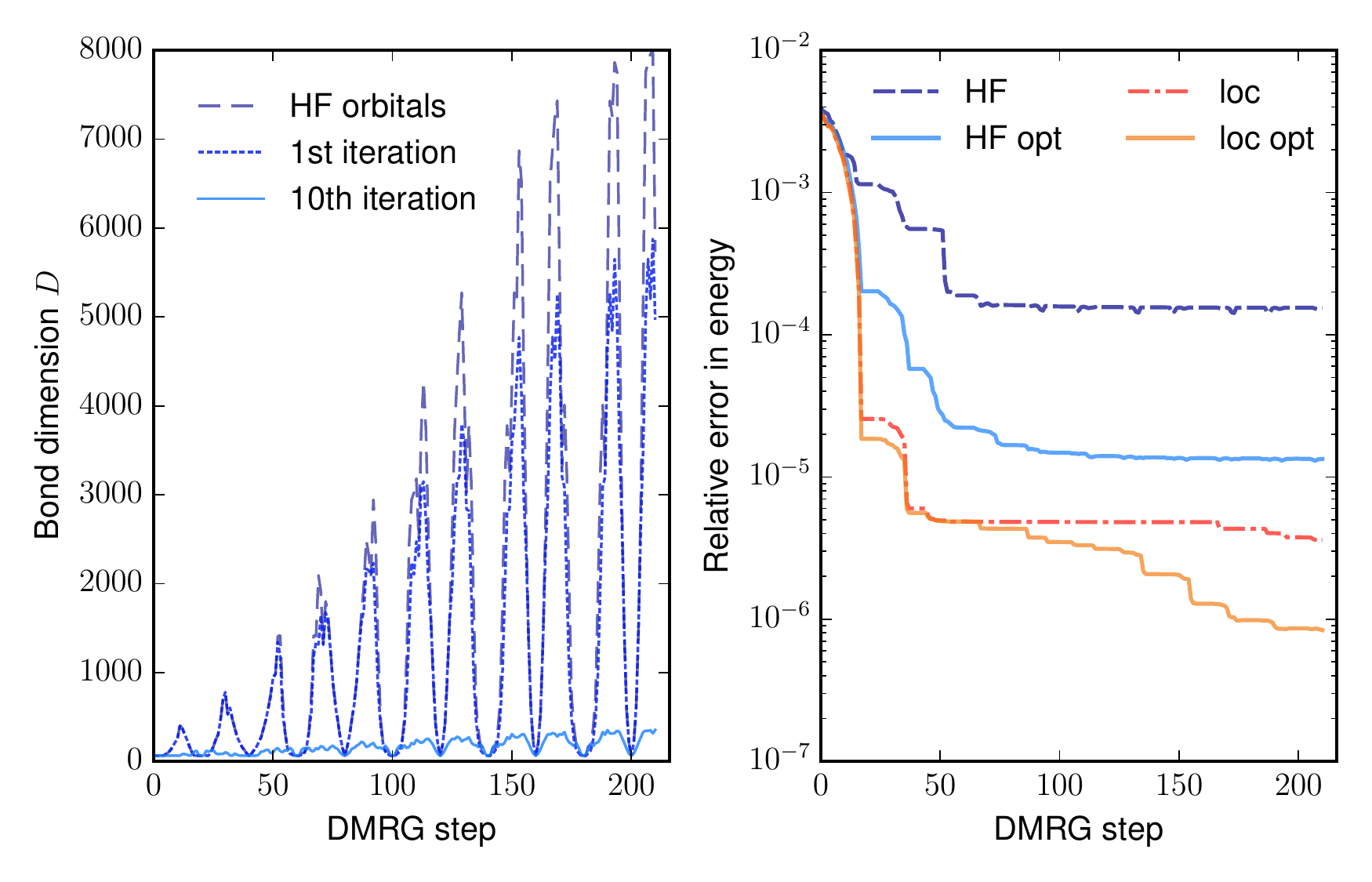}
	\caption{Numerical results for the Be ring of $6$ Be atoms with interatomic distance of $3.3$\,\AA. 
	All calculations have been performed with a $U(1)\times U(1)$ symmetric open boundary MPS and local mode transformations which keep the $SU(2)$ symmetry of the Hamiltonian. Both diagrams show results of the described optimisation for the physical basis.
	In the left panel we show the bond dimension needed for a bounded truncation error $\epsilon_{\rm trc}\leq 10^{-6}$ and $D_\text{min}=64$ when starting in the HF basis. 
	The dark blue dashed line corresponds to a calculation in the HF basis, the blue dotted and light blue line correspond to the first and the tenth iteration of the calculation with basis optimisation. 
	The right panel compares the relative error in energy $(\sandwich{\psi}{H}{\psi}-E_0)/E_0$ obtained by calculations with $D_\text{max}=256$, where the reference value for the ground state energy $E_0$ has been obtained from a calculation with $D_\text{max} = 2048$ in the localised basis.
	The dark blue dashed and red dashed-dotted line show the results for a calculation in the HF and localised basis respectively.
	In light blue and orange we plot the relative error of the 15th and 10th iteration of the calculation with basis optimisation starting in the HF and localised basis respectively.
	 }
	\label{fig:results}
\end{figure}

Standard QC-DMRG algorithms use complementary operators \cite{White1999,Chan2002,Legeza-2003a} in order to reduce the computational cost of each DMRG step. 
In the appendix we show that complementary operators transform as general operators under local mode transformations and argue that local mode transformation can be found and applied in a time not exceeding the computational cost of a single DMRG step.
This allows us to keep the structure and the computational complexity of the two-site DMRG algorithm and perform basis optimisations essentially for free with the algorithm in Table \ref{tbl:algorithm}.

{\it Numerical results.}
We use a QC-DMRG algorithm which uses the dynamical block state selection approach \cite{Legeza-2003a} and
configuration interaction based dynamically extended active space \cite{Legeza-2003b} procedure to accelerate the convergence and
adapt the basis of the physical space by the algorithm described above.
As a test-system, we have chosen
the electron-configuration of a Beryllium ring built from 6 Be atoms.
This system has recently been  investigated \cite{Fertitta-2014}
and a strong dependence of the convergence of the DMRG from the initial basis was observed.
We investigate the molecule in a stretched geometry with an interatomic distance of 3.3\,\AA.
As initial bases we use the Hartree-Fock (HF) basis of the system and a localised basis derived from the HF orbitals by a Foster-Boys localisation \cite{Boys-1960}.
Such localised orbitals are widely used in QC-DMRG calculations and are known to yield a better convergence for the Be ring \cite{Fertitta-2014}.

Starting from the according initial basis we iteratively apply the following scheme: 
we run the standard QC-DMRG for 2 sweeps, 
perform 8 additional sweeps together with the local mode transformation as described in Table \ref{tbl:algorithm} 
and reorder the basis according to its mutual information patterns \cite{Barcza-2011}.
Hereby we either fix the truncation error $\epsilon_{\rm trc}$ made in each step or set a hard-cut on $D_\text{max}$.
The results of our calculations are show in Fig.\ \ref{fig:results}. 
In the left panel of Fig.\ \ref{fig:results} we show how the bond dimension behaves for a ground state search 
with a bounded truncation error $\epsilon_{\rm trc}\leq 10^{-6}$ and $D_\text{min}=64$ 
for a calculation in the HF basis and optimised bases obtained by the above scheme starting from the HF orbitals. 

It is key to the method proposed that the optimisation of the basis leads to a significant decrease of needed resources already in the first iteration, 
where after the tenth iterations of basis optimisation the needed bond dimension is more than one order of magnitude smaller than in the unoptimised orbitals.
For realistic applications of the scheme intermediate high bond dimensions during the calculation need to be avoided.
The right panel of Fig.\ \ref{fig:results} shows the relative error in energy reached when performing a calculation with $D_\text{max}=256$ starting in the  HF and localised basis.
As noted before the localised basis allows for a more efficient approximation of the ground state than the HF basis.
The basis optimisation allows us to  further significantly optimise both the HF and the localised basis. 
Starting from the HF orbitals we obtain a basis for which the relative error in energy drops by one order of magnitude from $1.52\times 10^{-4}$ to $1.2\times10^{-5}$.
Beginning at the localised basis allows us to reduce the relative error in energy from $3.7\times10^{-6}$ to $8.3\times10^{-7}$.
Note that the energy in the optimised basis starting from the HF orbitals is slightly worse than the energy obtained in the localised orbitals, 
reflecting the fact that finding the optimal basis is a hard global optimisation problem.

We have repeated similar calculations for different configurations of the Be ring; at the equilibrium configuration and close to the avoided crossing.  
In each case we have been able to find a physical basis allowing for a more efficient approximation of the ground state.
This illustrates that the above scheme can significantly and efficiently optimise a given initial basis. 
As the local mode transformation can be added with no increase of the computational cost to an existing two-site DMRG and typically yield already 
in the first iteration of the basis optimisation a significant improvement of the basis our scheme extends the standard QC two-site DMRG.

{\it Conclusion and perspectives.}
In this work, we have presented a scheme that adapts the physical basis an MPS is formulated in by applying Gaussian transformations. 
Incorporating local Gaussian transformations into the two-site DMRG algorithm allows us to optimise both the basis and the MPS iteratively. 
The resulting algorithm successfully optimises the physical basis such that distinctly better approximations of the ground state of a given system by an MPS can be identified.

It should be manifest from the description of the method that the same idea
is equally applicable to other tensor networks, due to the locality of the 
transformations.
In particular, \emph{tree-tensor network} approaches \cite{Murg-2010a,Nakatani-2013,Murg-2015}
can readily be combined with the methods laid out here. Similarly, they are
expected to be helpful for 2-d lattice systems \cite{Verstraete-2004a,Verstraete-2008,Orus-2014}.
In addition the above scheme can be directly combined with recent developments for the time-evolution of MPS \cite{Haegeman-2014b} 
in order to obtain a time-evolution with variational physical basis.
 
Our general strategy -- of combining tensor networks with fermionic transformations -- complements the 
recent interesting approach of Ref.\ \cite{FermionicPollmann}, which is similar in mindset, but
where these two components are put together in the opposite order.
There, a matrix-product operator is applied onto a free fermionic wave function. In contrast to that
approach, we here retain efficient contractibility, however. The approach taken in this work can also be
seen as a variational principle that allows to find the optimal \emph{fermionic tensor network} in 
Ref.\ \cite{Ferris}, where a fixed fermionic basis change is being made use of.
Widening the scope, these tools seem also helpful in related approaches making 
use of a \emph{big data} machinery to capture strongly correlated quantum systems. For example, 
\emph{compressed sensing ideas} can help finding localised Wannier functions \cite{CS2,CS1}, 
which in turn can be made use of in density functional theory \cite{KohnSham,Scheffler}. 
In conjunction with the tools
developed here, a combined approach close to optimally representing fermionic
correlated states seems within reach.

{\it Acknowledgments.}
We would like to thank G.\ Chan, 
E.\ Fertitta, 
V.\ Murg, 
A.\ Nagy, 
V.\ Nebendahl, 
J.\ Rodriguez-Laguna,
R.\ Schneider,
T.\ Szilv\'asi, and
F.\ Verstraete
for discussions. J.\ E.\ and C.\ K.\ acknowledge support from the EU (SIQS, RAQUEL, AQUS), the ERC (TAQ),
the DFG (EI 519/7-1, CRC 183), the Templeton Foundation, and the Studienstiftung des Deutschen Volkes, 
L.\ V. and \"O.\ L.\ from the Czech Science Foundation, Grant No.\ 16-12052S and the Hungarian Research Fund (OTKA), Grant No.\ NN110360 and K100908.


%

\section*{Appendix}
\subsection*{Gradient and geometry of the optimisation problem}
\label{sec:App:Gradient}
\begin{figure}
	\includegraphics{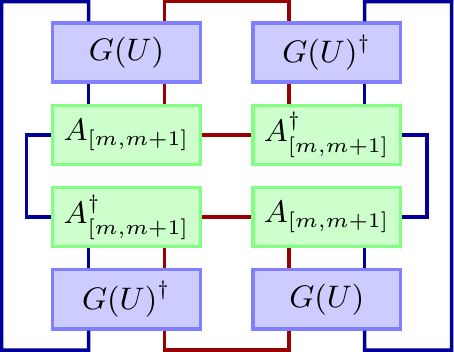}
	\caption{Tensor network representing $||\Sigma_{\psi(\id_m\oplus U\oplus\id_{n-m-2})}^m||^4_4$ with $U\in U(2p)$ if $\ket{\psi}$ is represented by a MPS in mixed normalised form as discussed in the main text with $A_{[m,m+1]}$ being the blocked tensor of sites $m$ and $m+1$.
	Legs corresponding to indices of site $m$ are indicated in blue, to indices of site $m+1$ in red.}
	\label{fig:4Norm}
\end{figure}

Implementing the optimisation methods in Grassmann manifolds as described in Refs.\ 
\cite{Edelman1998,Manton2002} we parametrise the Grassmann manifold $G(a,b)$ by $a\times b$ isometries $X\in \cc^{a\times b}$ which form the Stiefel manifold $V(a,b)$.
In order to implement a the conjugated gradient search for identifying a (local) minimum of $f:G(a,b)\rightarrow \rr$ in $G(a,b)$ the 
derivatives ${\partial f(X)}/{\partial \re X_{i,j}}+i {\partial f(X)}/{\partial \im X_{i,j}}$ are needed.
As the resulting mode transformation acts on two sites only, we are interested in the case $b={a}/{2}$.

Given an isometry $X\in \cc^{a\times a/2}$, i.e.$~X^\dag X =\id_{\frac{a}{2}}$ we can construct a unitary $U(X)\in U(n)$ with the first ${a}/{2}$ columns being equal to the columns of $X$ by
\begin{equation}
	U(X) = \id_a - (X-P)(\id_{\frac{a}{2}}-X^\dag P)^{-1} (X^\dag - P^\dag),
\end{equation}
with $P\in \cc^{a\times a/2}$ and $P_{i,j} = \delta_{i,j}$ which corresponds to a generalised Householder reflection of the subspace spanned by the columns of $X$. 
Note that if $\id_{\frac{a}{2}}-X^\dag P$ turns out to be singular, we can always transform the columns of $X$ using a random $a/2\times a/2$ unitary, e.g.~$e^{i\phi}\id_{\frac{a}{2}}$ with $\phi$ random, by which we chose a new representative in $V(a,a/2)$ of the same element in $G(a,a/2)$

Given a general invertible matrix $Y$, the elements of which depend on a real parameter $t$, we can evaluate the derivative of the inverse matrix to 
\begin{equation}
\frac{d}{dt}Y(t)^{-1} = Y(t)^{-1} [\frac{d}{dt} Y(t)] Y(t)^{-1}.
\end{equation}
From Fig.~\ref{fig:4Norm} we can read off 
\begin{eqnarray}
	&& ||\Sigma_\psi(\id_m\oplus U \oplus \id_{n-m-2})^m||_4^4 \nonumber\\
	&&=\tr([g(U)^\dag\otimes g(U)^\dag] M [g(U)\otimes g(U)] N), 
\end{eqnarray}
where $M(A_{[m,m+1]})$ corresponds to the inner ring of tensors in Fig.~\ref{fig:4Norm} which depends on the coupled tensor $A_{[m,m+1]}$ and $N$ orders the outer legs as shown. 
We can then evaluate the derivative of this cost function with respect to the parameters $\re(X_{i,j})$ and $\im(X_{i,j})$
using 
\begin{equation}
	\frac{\partial g(U)_{I,J}}{\partial U_{i,j}^\ast} = (-1)^{p_I(i)+p_J(j)}\det U^\dag|_{I\backslash\{i\},J\backslash\{j\}}
\end{equation}
if $|I|=|J|$, $i\in I$ and $j\in J$ and 0 otherwise, where $U^\dag|_{I,J}=(U^\dag_{i,j})_{i\in I,j\in J} $, $I,J\subset[np]$ and $p_X(x)$ denotes the number of elements of $X$ smaller $x$ and 
\begin{equation}
	\frac{\partial U(X(\alpha))}{\partial \alpha} = -X^\prime Z_2 - Z_1 X^\dag{}^\prime + Z_1 X^\dag{}^\prime P Z_2,
\end{equation}
with 
\begin{eqnarray}
	X^\prime_{i,j} &=& \frac{d X(\alpha)_{i,j}}{d \alpha},\\
	Z_1 &=& (X - P)(\id_{\frac{a}{2}}-X^\dag P)^{-1},\\
	 Z_2 &=& (\id_{\frac{a}{2}}-X^\dag P)^{-1} (X^\dag - P^\dag).
\end{eqnarray}

\subsection*{Preservation of the DMRG structure}
To avoid redundant calculations, DMRG implementations use complementary operators which are expanded and reloaded during the sweeps \cite{White1999,Chan2002,Szalay-2015}.
We denote by $L_m$ the set of all modes which are associated to sites $q$ with $q<m$ and by $R_m$ the set of modes that belong to sites $q$ with $q>m+1$. 
In addition we define the abbreviations 
\begin{equation}
	\ket{L_a} = \sum\limits_{\alpha_1,\ldots,\alpha_{m-1}=1}^d(A_{[1]}^{\alpha_1}\ldots A_{[m-1]}^{\alpha_{m-1}})_a\ket{\alpha_1}\otimes\ldots\otimes\ket{\alpha_{m-1}} 
\end{equation}	
and 
\begin{equation}\ket{R_a} = \sum\limits_{\alpha_{m+1},\ldots,\alpha_{n}=1}^d(A_{[m+1]}^{\alpha_{m+1}}\ldots A_{[n]}^{\alpha_{n}})_a\ket{\alpha_{m+1}}\otimes\ldots\otimes\ket{\alpha_{n}}.
\end{equation}	
The four different types of complementary operators, $P^{I_m,a,b}_{i,j},Q^{I_m,a,b}_{i,k},R^{I_m,a,b}_{i},S^{I_m,a,b}_{i}$ with $I_m=L_m,R_m$, $a,b\in[D_{m-1}]$ and $i,j,k\in [np]\backslash\{I_m\}$ are defined as follows
\begin{align}
	P^{L_m,a,b}_{i,j} &= \tr_{[m-1]}\Bigl[(\ketbra{L_a}{L_b}\otimes \id)\sum\limits_{k,l\in L_m} v_{i,j,k,l} c_i^\dag c_j^\dag c_l c_k\Bigr],\\
	Q^{L_m,a,b}_{i,k} &= \tr_{[m-1]}\Bigl[(\ketbra{L_a}{L_b}\otimes \id)\sum\limits_{j,l\in L_m} v_{i,j,k,l} c_i^\dag c_j^\dag c_l c_k\Bigr],\\
	R^{L_m,a,b}_{i} &= \tr_{[m-1]}\Bigl[(\ketbra{L_a}{L_b}\otimes \id)\sum\limits_{j,k,l\in L_m} v_{i,j,k,l} c_i^\dag c_j^\dag c_l c_k\Bigr],\\
	S^{L_m,a,b}_{i} &= \tr_{[m-1]}\Bigl[(\ketbra{L_a}{L_b}\otimes \id)\sum\limits_{j\in L_m} t_{i,j} c^\dag_ic_j\Bigr],
\end{align}
where the partial traces are evaluated according to the Jordan Wigner representation, plus their corresponding counterparts on sites $q$ with $q>m+1$ which are defined analogously by replacing $L_m$ by $R_m$, $(\ketbra{L_a}{L_b}\otimes\id)$ by $(\id\otimes\ketbra{R_a}{R_b})$ the partial trace over sites $[m-1]$ by the partial trace over sites $[n]\backslash[m+1]$.
During a DMRG run the complementary operators for different $m$ are saved and reused in later steps.
In order to update the complementary operators efficiently without loading and saving many operators from and to the disk, we evaluated some of the basis changes in a lazy fashion. During a right sweep, the complementary operators $P^{R_m}$, $Q^{R_m}$, $R^{R_m}$ and $S^{R_m}$ are loaded while a mode transformation transforming the orbitals $1,\ldots,m-1$ was accumulated (resulting from the previous $2m-3$ steps).
We obtain the complementary operators within the updated basis by transforming them similar to general operators as discussed in the main text by 
\begin{align}
	P^{I_m,a,b}_{i,j}(U)&=\sum\limits_{i^\prime,j^\prime\in[np]\backslash I_m} (U_{I}^\dag)_{i,i^\prime} (U_{I}^\dag)_{j,j^\prime} P^{I_m,a,b}_{i^\prime,j^\prime}(\id)\nonumber\\
			&\:= [(U_I^\dag\otimes U_I^\dag) P^{L_m,a,b}(\id)]_{i,j} ,\\
	Q^{I_m,a,b}(U) &= U^\dag_{I}Q^{I_m,a,b}(\id) U_I , \\
	R^{I_m,a,b}(U) &= U_I^\dag R^{I_m,a,b}(\id) , \\
	S^{I_m,a,b}(U) &= U_I^\dag S^{I_m,a,b}(\id),
\end{align}
with $I=L,R$ for a left and right sweep respectively and $U_L$ and $U_R$ the corresponding accumulated transformations. Note that, for a right sweep, the operators $P^{L_m}$, $Q^{L_m}$, $R^{L_m}$ and $S^{L_m}$ can be formed with the rotated coefficients of the Hamiltonian and need no further rotation.

\subsection*{Cost-analysis for the presented scheme}
The run time for a two-site DMRG per DMRG-step including an update of the complementary operators when stepping from site $m$ to $m\pm1$ during a sweep as $\landau(n^2 D^3 2^{2p} + n^2 D^2 2^{3p} + n^3 D^2 p^2)$ \cite{Chan2002}.
The steps of the algorithm presented in Table 1 in the main text scale in addition to that as follows. Note that $p$ is typically 2 (spin up and down electrons) for a QC-DMRG - but we will keep track of the dependence of the cost on $p$ for completeness.
The rotation of the blocked matrix and computation of $f^{(1)}_m$ comes at a cost of $\landau(D^2 2^{4p} + D^32^{3p})$. 
The transformation of the conjugated operators are performed with a cost of $\landau(n^3 p^3 D^2)$ for general transformations, where the cost can be lowered to $\landau(n^3 p^2 D^2)$, once a $U(1)^{\times p}$ or more general symmetry is enforced on the transformations.
Due to the locality of the mode transformation, we can transform coefficients of the Hamiltonian in a time scaling as $\landau(n^3p^3)$.
The computational cost induced from the local mode transformations which optimise $f^1_m$ iteratively are therefore given by $\landau(D^2 2^{4p} + D^3 2^{3p} + n^3 p^3 D^2)$, so for the relevant parameters the number of orbitals $n$ and the bond dimension $D$ the additional cost scale as $\landau(D^3+n^3D^2)$ which is lower in cost than the two-site DMRG step with $\landau(n^2D^3 + n^3 D^2)$.

\subsection*{Two-site correlations in the different physical bases }

\begin{figure}[h]
 \includegraphics[width=0.4\textwidth]{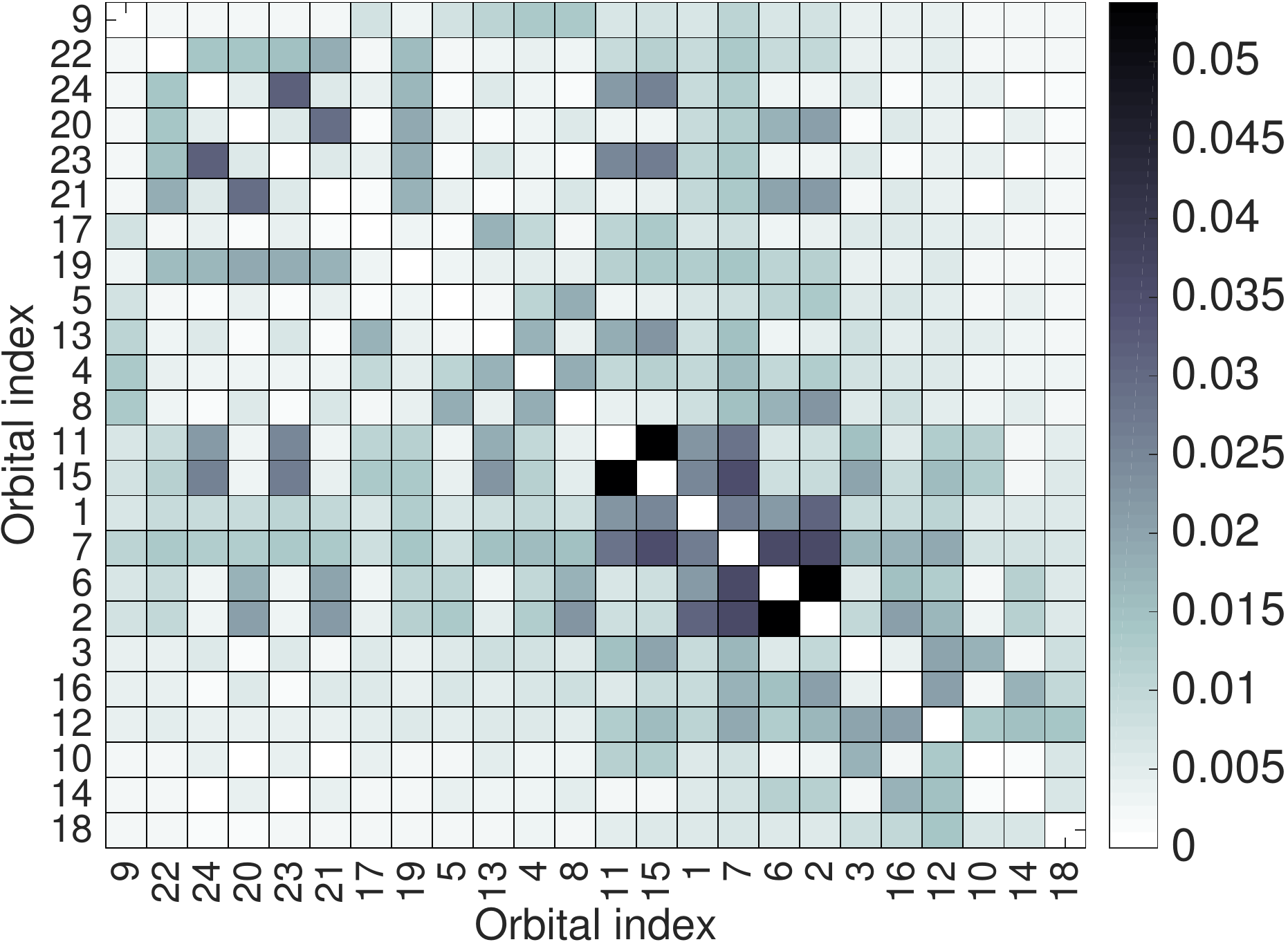}
	\caption{Mutual information present in the ground state approximation within the HF basis by an $U(1)\times U(1)$-symmetric MPS with $D_\text{max} = 256$ (dark blue dashed curve in the right panel of Fig.\ 2 in the main text).}
	\label{fig:MutIcan0}
\end{figure}

\begin{figure}[h]
 \includegraphics[width=0.4\textwidth]{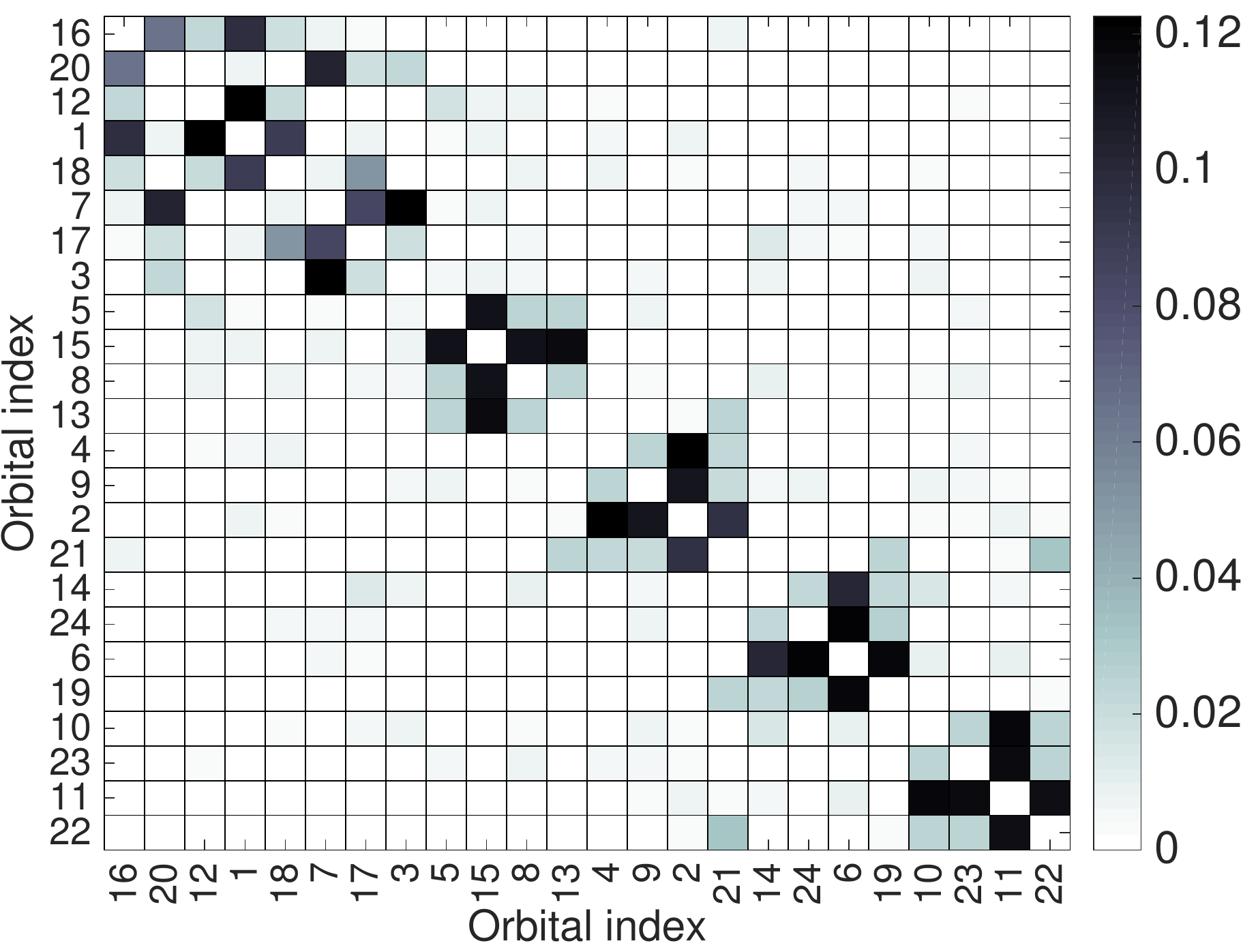}
	\caption{Mutual information present in the ground state approximation within the optimised HF basis by an $U(1)\times U(1)$-symmetric MPS with $D_\text{max} = 256$ (light blue curve in the right panel of Fig.\ 2 in the main text).}
	\label{fig:MutIcan15}
\end{figure}

\begin{figure}[h]
 \includegraphics[width=0.4\textwidth]{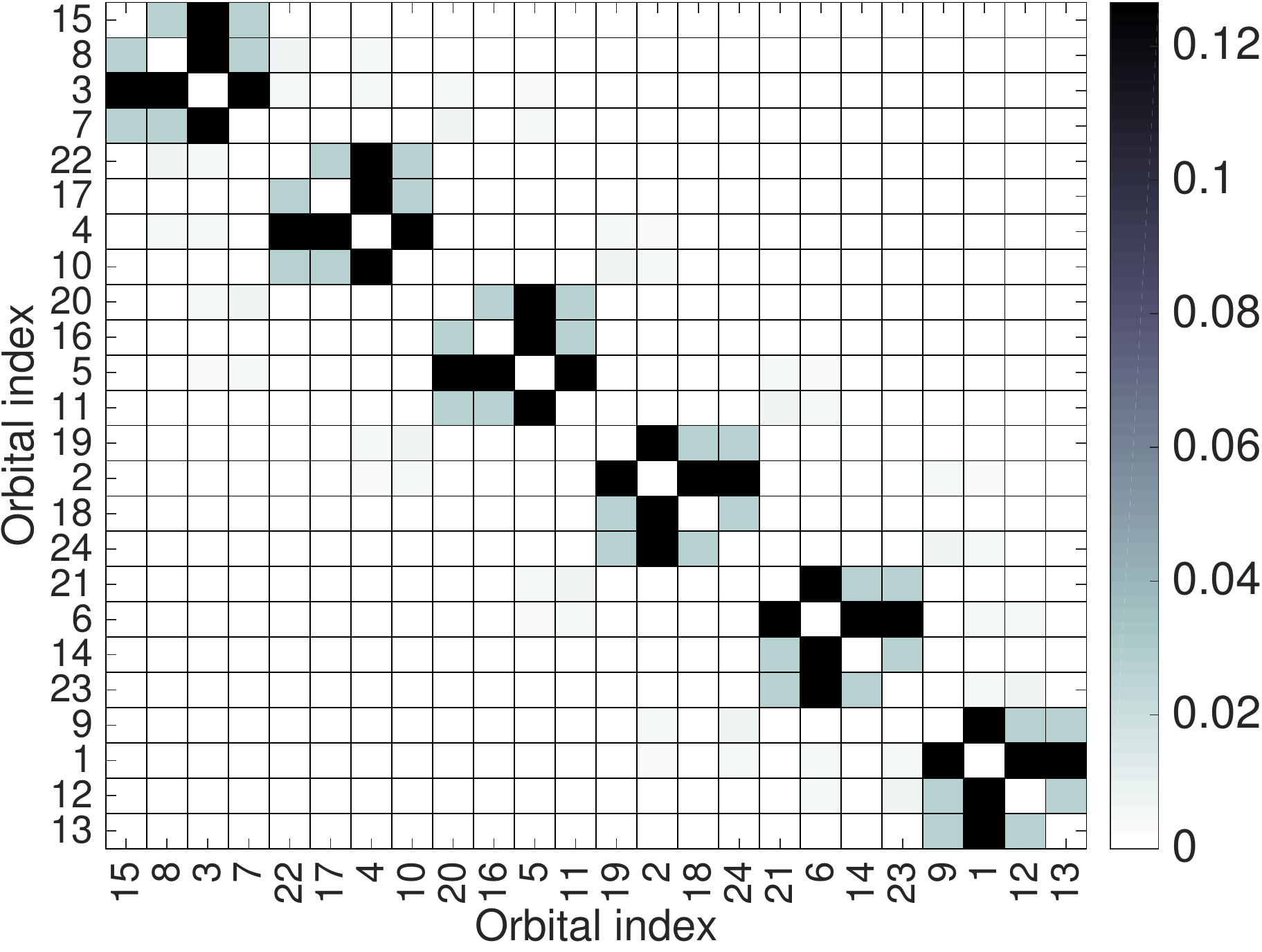}
	\caption{Mutual information present in the ground state approximation within the localised basis by an $U(1)\times U(1)$-symmetric MPS with $D_\text{max} = 256$ (red dashed-dotted curve in the right panel of Fig.\ 2 in the main text).}
	\label{fig:MutIloc0}
\end{figure}

\begin{figure}[h]
 \includegraphics[width=0.4\textwidth]{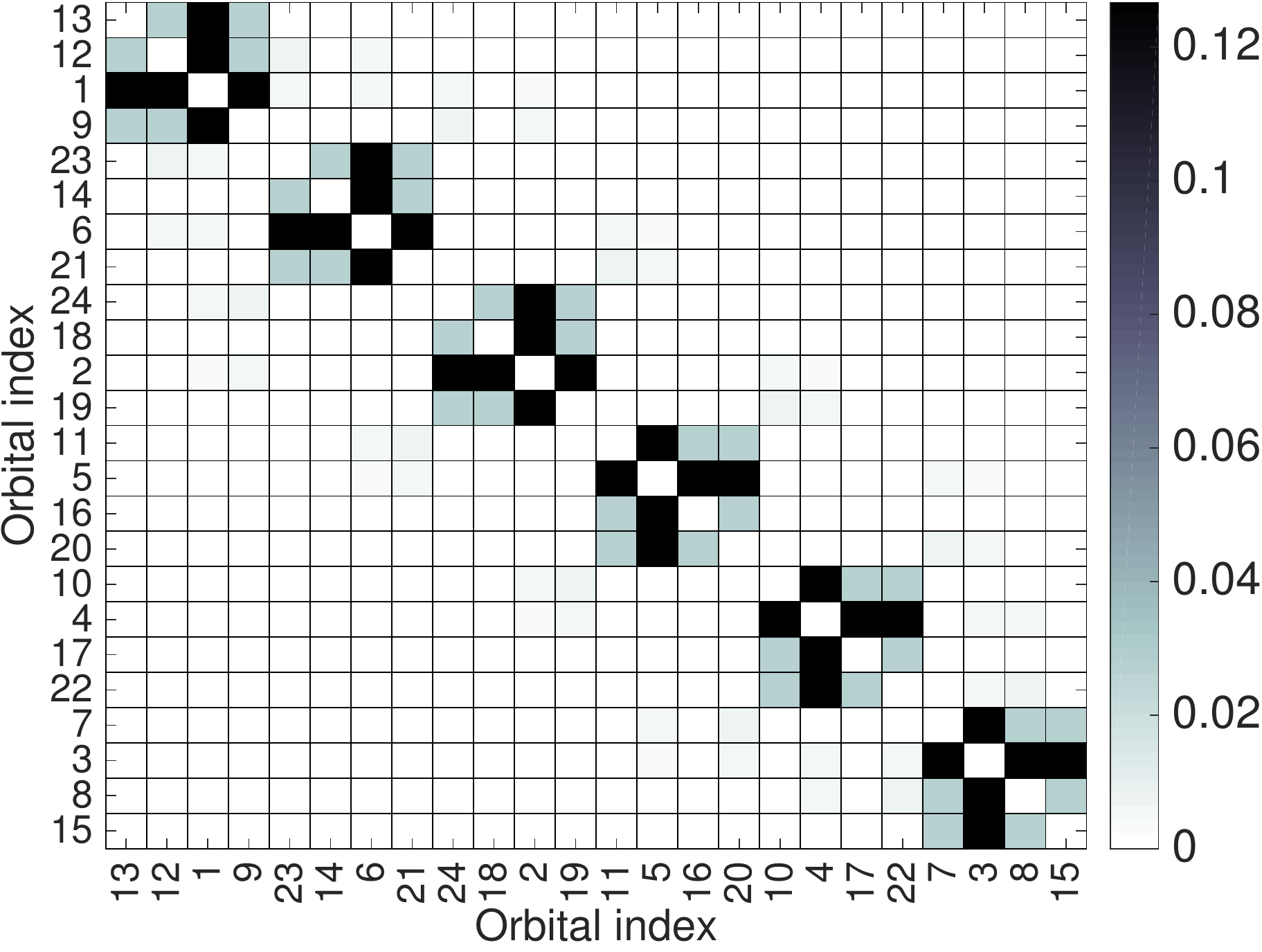}
	\caption{Mutual information present in the ground state approximation within the optimised localised basis by an $U(1)\times U(1)$-symmetric MPS with $D_\text{max} = 256$ (orange curve in the right panel of Fig.\ 2 in the main text).}
	\label{fig:MutIloc10}
\end{figure}

\begin{figure}[h]
 \includegraphics[width=0.4\textwidth]{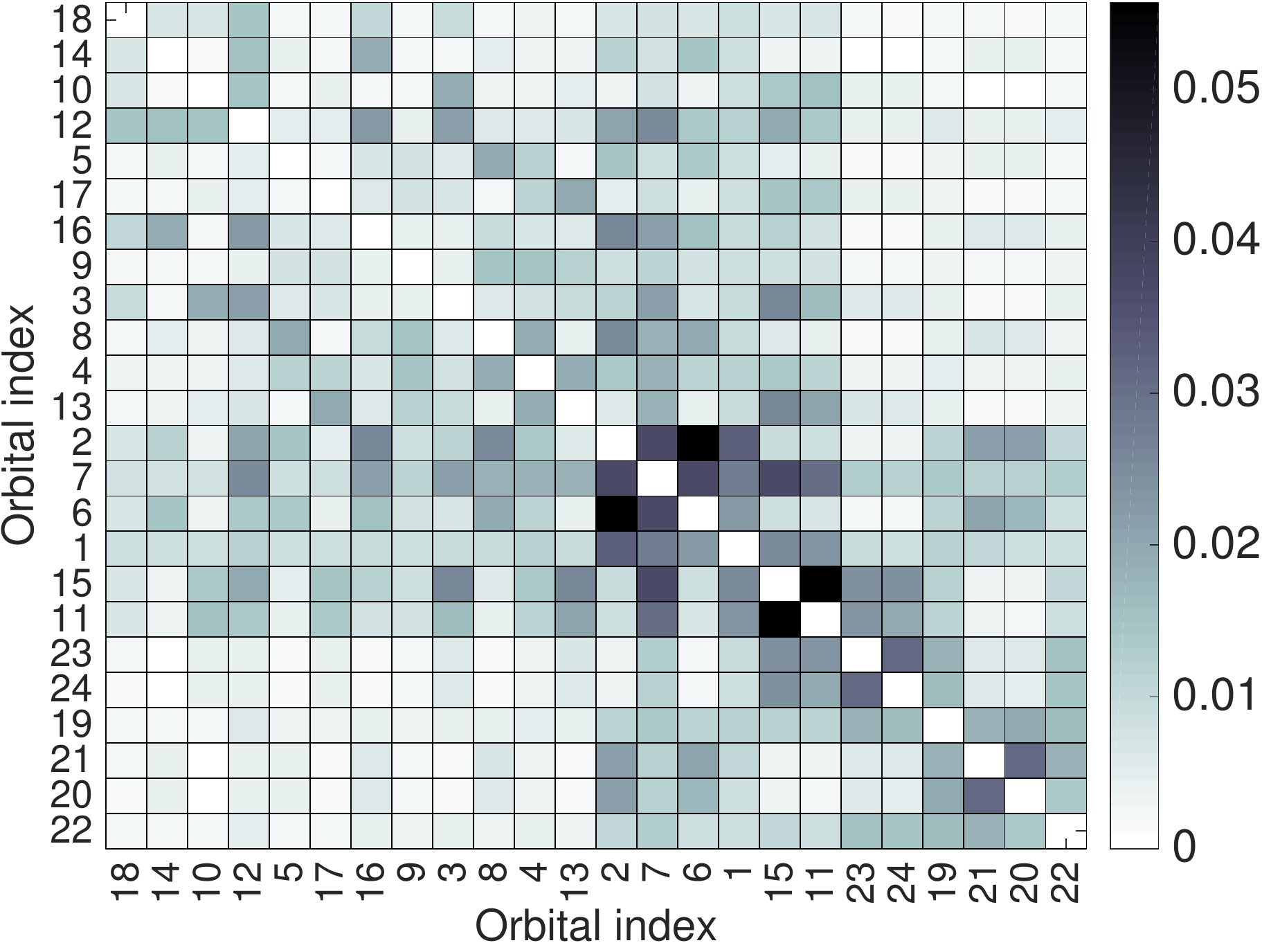}
	\caption{Mutual information present in the ground state approximation within the HF basis by an $U(1)\times U(1)$-symmetric MPS with a bounded truncation error $\epsilon_{\rm trc}\leq 10^{-6}$ (dark blue dashed curve in the left panel of Fig.\ 2 in the main text).}
	\label{fig:MutIetr0}
\end{figure}

\begin{figure}[h]
 \includegraphics[width=0.4\textwidth]{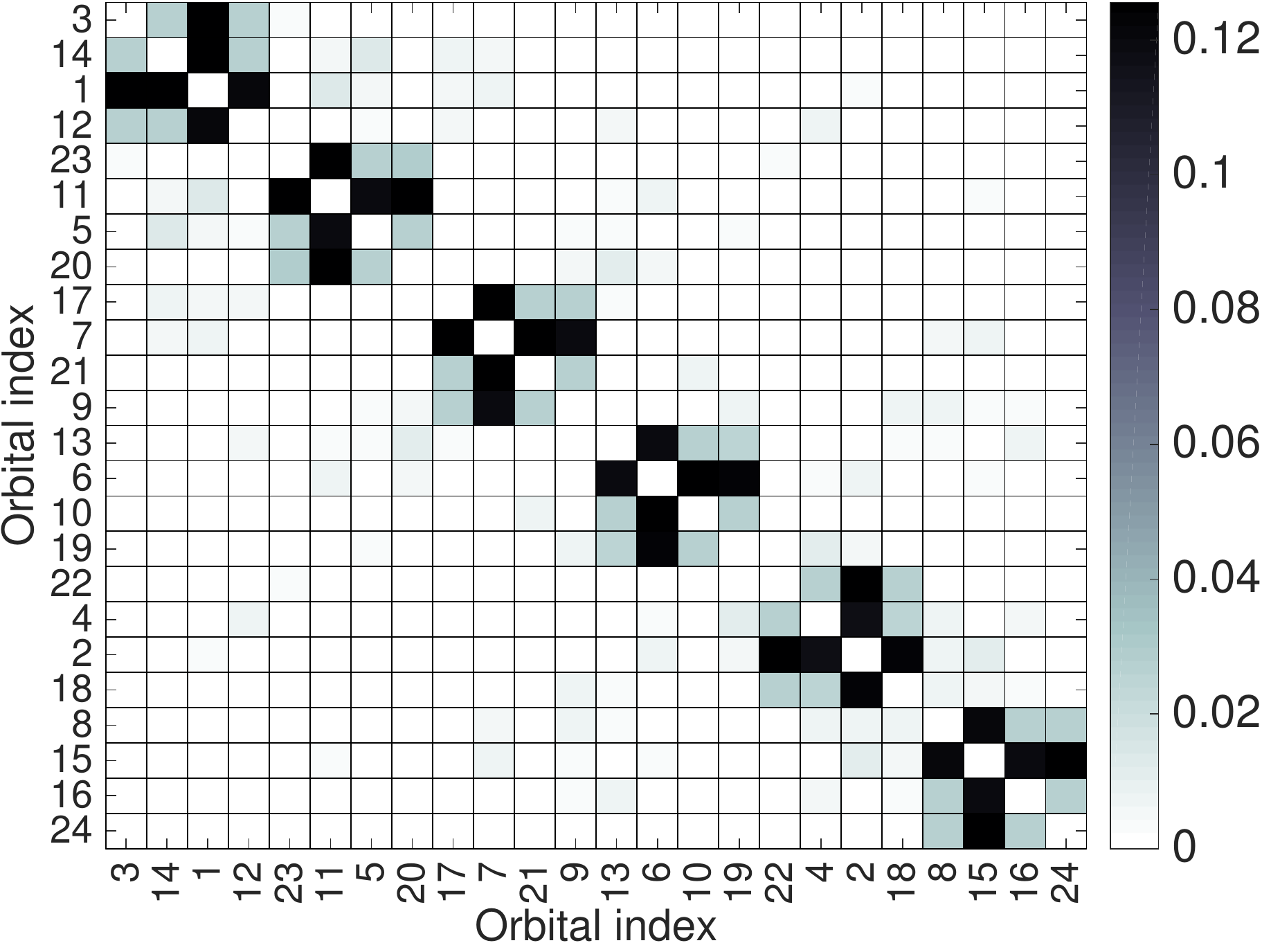}
	\caption{Mutual information present in the ground state approximation after the tenth basis optimisation by an $U(1)\times U(1)$-symmetric MPS with a bounded truncation error $\epsilon_{\rm trc}\leq 10^{-6}$ (light blue curve in the left panel of Fig.\ 2 in the main text).}
	\label{fig:MutIetr10}
\end{figure}

In order to visualize the different physical bases we present below plots for the two site correlations present in the ground state approximations within the corresponding basis.
The two site correlation is measured here by the mutual information $I(q,r) = S^1(q) + S^1(r) - S^1(q,r)$ with $q,r\in[n]$ and $S^1(I)$ denotes the von Neumann entropy of the reduced state $\rho_I = \tr_{[n]^\backslash I}|\psi\rangle\langle \psi |$. 
We present the correlation patterns of the final states of the individual calculations shown in Fig.\ 2 in the main text. Fig.\ \ref{fig:MutIcan0}--\ref{fig:MutIloc10} show the mutual information of the final states obtained at $D_\mathrm{max}=256$ in the initial and optimised basis starting in the HF and localised orbitals. Fig.\ \ref{fig:MutIetr0}, \ref{fig:MutIetr10} display the mutual information patterns obtained in the calculation with bounded truncation error $\epsilon_{\rm trc} = 10^{-6}$.

The correlation patterns obtained in the localised (Fig.\ \ref{fig:MutIloc0} ) and optimised bases (Fig.\ \ref{fig:MutIcan15}, \ref{fig:MutIloc10}, \ref{fig:MutIetr10}) show very similar features, highlighting the fact that the localised basis in this system will be close to optimal. Note, however, that although the correlations for the localised and optimised version of the localised basis are almost the same (compare Fig.\ \ref{fig:MutIloc0} and Fig.\ \ref{fig:MutIloc10}), the latter allows for a more efficient approximation of the ground state using a QC-DMRG; displaying the fact that the mutual information does not contain the full information about the optimality of the basis.

\end{document}